\documentclass{iau}
\usepackage{graphicx}
\usepackage{amssymb}

\def\apj {ApJ}

\def\aj {AJ}
\def\mnras {MNRAS}

\def\araa {ARAA}

\def\bea{\begin{eqnarray}}
\def\eea{\end{eqnarray}}
\def\bee{\begin{equation}}
\def\eee{\end{equation}}
\def\bef{\begin{figure}}
\def\eef{\end{figure}}
\def\befs{\begin{figure*}}
\def\eefs{\end{figure*}}
\def\be{\begin{equation}}
\def\ee{\end{equation}}

\def\ff   {\ifmmode{f}\else{$f$}\fi}
\def\rmax {\ifmmode{r_{\rm max}}\else{$r_{\rm max}$}\fi}
\def\zmax {\ifmmode{z_{\rm max}}\else{$z_{\rm max}$}\fi}

\def\fCNM {\ifmmode{f^{\rm CNM}}\else{$f^{\rm CNM}$}\fi}
\def\fCMM {\ifmmode{f^{\rm CMM}}\else{$f^{\rm CMM}$}\fi}
\def\fWNM {\ifmmode{f^{\rm WNM}}\else{$f^{\rm WNM}$}\fi}
\def\fWIM {\ifmmode{f^{\rm WIM}}\else{$f^{\rm WIM}$}\fi}
\def\fCM  {\ifmmode{f^{\rm CM}} \else{$f^{\rm CM}$}\fi}
\def\fWM  {\ifmmode{f^{\rm WM}} \else{$f^{\rm WM}$}\fi}

\def\noCNM {\ifmmode{n_o^{\rm CNM}}\else{$n_o^{\rm CNM}$}\fi}
\def\noCMM {\ifmmode{n_o^{\rm CMM}}\else{$n_o^{\rm CMM}$}\fi}
\def\noWNM {\ifmmode{n_o^{\rm WNM}}\else{$n_o^{\rm WNM}$}\fi}
\def\noWIM {\ifmmode{n_o^{\rm WIM}}\else{$n_o^{\rm WIM}$}\fi}
\def\noCM  {\ifmmode{n_o^{\rm CM}} \else{$n_o^{\rm CM}$}\fi}
\def\noWM  {\ifmmode{n_o^{\rm WM}} \else{$n_o^{\rm WM}$}\fi}

\def\nnCNM {\ifmmode{n^{\rm CNM}}\else{$n^{\rm CNM}$}\fi}
\def\nnCMM {\ifmmode{n^{\rm CMM}}\else{$n^{\rm CMM}$}\fi}
\def\nnWNM {\ifmmode{n^{\rm WNM}}\else{$n^{\rm WNM}$}\fi}
\def\nnWIM {\ifmmode{n^{\rm WIM}}\else{$n^{\rm WIM}$}\fi}
\def\nnCM  {\ifmmode{n^{\rm CM}} \else{$n^{\rm CM}$}\fi}
\def\nnWM  {\ifmmode{n^{\rm WM}} \else{$n^{\rm WM}$}\fi}

\def\zCNM {\ifmmode{z_o^{\rm CNM}}\else{$z_o^{\rm CNM}$}\fi}
\def\zCMM {\ifmmode{z_o^{\rm CMM}}\else{$z_o^{\rm CMM}$}\fi}
\def\zWNM {\ifmmode{z_o^{\rm WNM}}\else{$z_o^{\rm WNM}$}\fi}
\def\zWIM {\ifmmode{z_o^{\rm WIM}}\else{$z_o^{\rm WIM}$}\fi}
\def\zCM  {\ifmmode{z_o^{\rm CM}} \else{$z_o^{\rm CM}$}\fi}
\def\zWM  {\ifmmode{z_o^{\rm WM}} \else{$z_o^{\rm WM}$}\fi}

\def\nH {\ifmmode{\langle n_H \rangle}\else{$\langle n_H \rangle$}\fi}
\def\Em {\ifmmode{E_m}\else{$E_m$}\fi}
\def\deg   {\ifmmode{^\circ}\else{$^\circ$}\fi}
\def\HH    {\ifmmode{\cal H}\else{${\cal H}$}\fi}
\def\dH    {\ifmmode{\delta\cal H}\else{$\delta{\cal H}$}\fi}
\def\Hdisk {\ifmmode{\cal H}_{\rm disk}\else{${\cal H}_{\rm disk}$}\fi}
\def\Hdiskp {\ifmmode{\cal H}^{'}_{\rm disk}\else{${\cal H}^{'}_{\rm disk}$}\fi}
\def\Hsky  {\ifmmode{\cal H}_{\rm sky}\else{${\cal H}_{\rm sky}$}\fi}

\def\gta{\;\lower 0.5ex\hbox{$\buildrel > \over \sim\ $}}
\def\lta{\;\lower 0.5ex\hbox{$\buildrel < \over \sim\ $}}          
\def \NH {\ifmmode{\rm N}_{\scriptscriptstyle H}\else{N$_{\scriptscriptstyle H}$}\fi}

\def\deg{\hbox{${}^\circ$}}

\def\las{\mathrel{\hbox{\rlap{\hbox{\lower4pt
        \hbox{$\sim$}}}\hbox{$<$}}}}
\def\gas{\mathrel{\hbox{\rlap{\hbox{\lower4pt
            \hbox{$\sim$}}}\hbox{$>$}}}}

\def\baryfrac{$\langle f_b \rangle$}

\def\fig#1{Fig.~\ref{fig:#1}} 

\def\equ#1{Equation~(\ref{equ:#1})}

\title[A mechanism for galaxy bulk flows] 
{Void asymmetries in the cosmic web: a mechanism for bulk flows}

\author[J. Bland-Hawthorn \& S. Sharma]   
{J. Bland-Hawthorn$^1$
 \and S. Sharma$^1$}

\affiliation{$^1$Sydney Institute for Astronomy, University of Sydney, \\
School of Physics A28, NSW 2006, Australia \\
email: {\tt jbh@physics.usyd.edu.au}}

\pubyear{2014}
\volume{308}  
\pagerange{}
\jname{The Zel'dovich Universe}
\editors{R. van der Weygaert eds.}
\begin{document}

\maketitle

\begin{abstract}
Bulk flows of galaxies moving with respect to the cosmic microwave
background are well established observationally and seen in the most 
recent $\Lambda$CDM simulations. With the aid of an idealised Gadget-2
simulation, we show that void asymmetries in the cosmic web can
exacerbate local bulk flows of galaxies.
The {\it Cosmicflows-2} survey, which has mapped in 
detail the 3D structure of the Local Universe, reveals that the Local Group 
resides in a ``local sheet'' of galaxies that
borders a ``local void'' with a diameter of about 40 Mpc. The void
is emptying out at a rate of 16 km s$^{-1}$ Mpc$^{-1}$.
In a co-moving frame, the Local Sheet is found to be moving away from
the Local Void at $\sim 260$ km s$^{-1}$. Our model
shows how asymmetric collapse due to unbalanced voids on either side
of a developing sheet or wall can lead to a systematic movement of the
sheet. We conjectured that asymmetries could lead to a large-scale 
separation of dark matter and baryons, thereby driving a dependence of 
galaxy properties with environment, but we do {\it not} find any evidence 
for this effect.
\end{abstract}


\firstsection 
              
\section{Introduction}
The physics of baryons across the universe is the grandest of all
environmental sciences. This drama is played out against a backdrop of
evolving dark matter structure from the 
Big Bang to the present day. Cold dark matter (CDM) simulations without baryons 
reveal a universe that looks structurally different from the observed universe 
defined by its baryons. The most recent CDM simulations that
include baryons and hydrodynamics emphasise how little we know of 
baryonic processes throughout cosmic time (Schaye et al 2014; Vogelsberger et al 2014). 

This meeting honours the 100th year since the birth of Y.B. Zel'dovich. In 1977,
Tallinn, Estonia was the site of the first great conference on large-scale structure. 
Over the past week, much of the discussion centred on where next for studies of
the cosmic web and galaxy redshift surveys. An interesting question
is how the ratio of baryons to dark matter by mass (\baryfrac)
varies across large-scale structure.  Non-standard models do
exist which predict baryon to dark matter variations (Malaney \& Mathews 1993;
Gordon \& Lewis 2003). 
Variations in \baryfrac\ of order 10\% lead to only few percent variations in the 
matter power spectrum, but could conceivably lead to observable
differences in some local galaxy properties (Nichols \& Bland-Hawthorn 2013).
In clusters, the baryon fraction 
approaches the universal average $\langle f^o_b \rangle \approx 15.5\%$
(Planck) with small scatter (Sun et al 2009).
For most galaxies in groups, this ratio is more uncertain largely because 
the warm-hot gas phases are very difficult to detect. In some instances, the
majority of the missing baryons may be in a warm circumgalactic medium 
(Tumlinson et al 2011; Shull et al 2012).

The next generation of large-scale galaxy surveys will seek to associate
more of a galaxy's properties with its large-scale environment (Croom et al 2012;
Bundy et al 2014; Bland-Hawthorn 2014). While the
distinction between clusters and the field is well defined, the dependence
of a galaxy's properties on a more graded local density has been hard to 
establish (e.g. Blanton \& Moustakas 2009; Metuki et al 2014). The effects appear to exist
only weakly, if at all. These include a weak dependence of the fundamental
plane with environment, scatter in the mass-metallicity relation that correlates
with environment (Cooper et al 2008), and mean star formation rates showing
a trend with environment (Lewis et al 2002; Gomez et al 2003).
The weakness of these trends may arise from any of the following: 
(i) the difficulty of defining environment;
(ii) the wrong galaxy parameters are being explored; 
(iii) a strong local dependence does not exist in nature.

In light of recent simulations where void-void imbalances are observed to push material
around (Pichon et al 2011; Codis et al 2012), we look more closely at the prospect
of baryon-dark matter variations. While the effects look strong in our 1D toy model, they are 
essentially non-existent in 3D. But what we do find is a barycentric drift of the collapsed
sheet and a possible mechanism for bulk flows in galaxies (e.g. Rubin et al 1976; Burstein
et al 1990; Tully et al 2008).

In \S 2, we introduce our 1D toy model for asymmetric collapse that suggests
a strong separation of dark matter and baryons. We investigate this idea further in \S 3
with a cosmologically motivated 3D model using Gadget-2.
In \S 4, we conclude that there is no general case for baryon-dark
matter separation on megaparsec scales, but we establish an interesting mechanism
for bulk flows in the presence of void asymmetries in the cosmic web.

\section{Toy model}

Can baryons and dark matter separate on megaparsec scales? The short
answer is yes in specific cases, for example, the Bullet Cluster (e.g. Mastropietro
\& Burkert 2008) where two massive clusters have passed through each 
other sweeping out the baryons in both systems. This provides us with our
initial motivation for modelling the Local Sheet. To illustrate how the Local Sheet
can be kinematically offset from the Local Void, initially, we reduce the
the dynamics of a forming sheet to a one-dimensional problem (cf. Melott 1983).
Before exploring a cosmologically motivated N-body simulation in
co-moving coordinates,
we develop a toy model using the ``infinite sheets'' approximation 
(Binney \& Tremaine 2008, hereafter BT08).
This approach has a long history dating back to theoretical work in plasma physics (Eldridge \& Feix 1963)
although its relevance to structure formation has been demonstrated in numerous papers (Yamashiro et al 1992). An 
up-to-date discussion is given by Teles et al (2011) who call for the sheets approximation
to be explored in a cosmological context, as we do here.

We simulate the formation and evolution of the Local Sheet using
thin sheets of dark matter and a matching set of (initially cospatial) sheets
made of gas. The dark matter is treated as 
collisionless while the gas is assumed to undergo inelastic collisions.
In the expanding universe, dark matter and baryons turn around and begin to 
collapse towards a local density perturbation. At turnaround, when no sheet 
crossing has occurred, the evolution is described by linear theory.
But during the collapse phase, the sheets start to cross each other,  
and the evolution becomes non-linear. 
We start the simulation just after turn around. Initially we treat the symmetric
case where the sheet separations have a Gaussian distribution in the normal ($x$-axis) 
direction. The gas and dark matter sheets are assumed to extend to infinity in the $y-z$ 
plane such that the force exerted by any sheet is constant at any point

The equation of motion for the sheets along the $x$ axis is given by
\bee
\ddot{x} = 2\pi G \int_{-\infty}^{\infty} {\rm Sgn}(x'-x)\sigma(x') dx'
\eee
or equivalently
\bee
\ddot{x} = f(x)= 2\pi G (2\Sigma(>x)-\Sigma_{\rm tot})
\label{eq:eq1}
\eee
where $\Sigma(>x)$ is the cumulative surface density and it is assumed that
$\Sigma(<x)+\Sigma(>x)=\Sigma_{\rm tot}$. For a discretized system, 
one can think of $2N+1$ sheets distributed in space with $i$-th sheet
having surface density $m_{i}$. $N$ must be large enough to render the
system ``collisionless'' as discussed by Yamashiro et al (1992).
The mass of the $i$-th sheet is assumed 
to be distributed evenly between $(x_{i-1}+x_{i})/2$ and
$(x_{i+1}+x_{i})/2$. Here we assume that all dark matter sheets have the
same constant surface density; the gas sheets also have a constant
surface density defined to be a factor of 10 lower (\baryfrac=0.1) than for 
the dark matter.
In our analysis, we set $G=1$ and choose 
$\Sigma_{\rm tot}=\int \sigma(x)dx=1/(2\pi)$.

\noindent
{\it Initial conditions.} Let $\sigma(\xi)$ ($-1<\xi<1$) be the initial density distribution at
shortly after the Big Bang ($t=0$).
If no shell crossings have happened since the Big Bang, the acceleration is
constant and is given by
\bee
a(\xi)=-2\pi G \int_{-1}^{1} {\rm Sgn}(\xi'-\xi)\sigma(\xi') d\xi'.
\eee 
Let $v(\xi)=v_{\rm tot} \xi$ be the initial velocity field.
The collapse time is given by 
\bee
t_{\rm collapse}=-\frac{2v(\xi=1)}{a(\xi=1)}=\frac{ v_{\rm  tot}}{\pi G \Sigma_{\rm tot}}
\eee 
The position and velocity at a later time $\tau$ is given by
\bee
x(\xi,\tau)=v_{\rm tot}\xi \tau+\frac{a(\xi)\tau^2}{2} \\
\ v(\xi,\tau)=v_{\rm tot}\xi+a(\xi)\tau .
\eee 
After BT08, we set $v_{\rm tot}=0.75$, which at $\tau=1$
gives max($v$)=max($x$)=0.25.

Let the initial sheet distribution be given by a function of form 
\be
\sigma(\xi)= \frac{k}{1-a {\rm cos}(b\pi\xi)} {\rm \ \ },
-1<\xi<1.
\label{equ:perturbation}
\ee
Using $\int_{-1}^{1} \sigma(\xi) d\xi=1$, the 
normalization constant is given by  
\bee
k=\frac{\pi b \sqrt{1-a^2}}{4{\rm tan^{-1}}((1+a) {\rm tan}(b\pi/2)/\sqrt{1-a^2})}
\eee
To sample such a distribution we use the method of inverse  
transform sampling. Let $F(>\xi)$ be the cumulative distribution, 
then for $u$ uniformly sampled between 0 and 1, the $\xi$ is given by  
\bea
\xi & = & F^{-1}(u) \\
    & = & \frac{2}{b\pi}{\rm tan}^{-1}\left[ \frac{\sqrt{1-a^2}}{1+a} {\rm tan}\left(\frac{(b\pi
      \sqrt{1-a^2})}{2k}(u-0.5)\right)\right]
\eea
Here we set $a=0.3$ and $b=0.3$ (BT08).

For the asymmetric case, the density is defined as follows
$$
\Sigma'(\xi) = \left\{ \begin{array}{rl}
  \Sigma(\xi) &\mbox{ if $\xi<0$} \\
  2 \Sigma(2\xi) &\mbox{ if $\xi>0$}
       \end{array} \right.
$$
This is achieved by setting $\xi=\xi/2$ for $\xi>0$. The position 
and velocities are calculated as in the symmetric case. 
Note for the region $\xi>0$, the collapse time now decreases by a factor of two.

To evolve a system of sheets, we use the kick-drift-kick
algorithm. For fixed time step $\Delta t$ this is given as
\bea
v_i & = & v_i+f(x_i)\Delta t \\
x_i & = & x_i+v_i \Delta t  \\
v_i & = & v_i+f(x_i)\Delta t
\eea
A time step of $\Delta t= 10^{-4} \tau$ was employed in all runs. 
The results were checked for convergence: choosing a lower 
time step did not yield any difference in results. 
For collisionless sheets, it is sufficient to just evolve the 
as shown above, but for gas one has to put in additional 
physics. We assume the gas sheets undergo fully inelastic collisions,
which conserve mass and momentum, but not energy. The prescription 
to simulate this is as follows. In a given time step, we identify 
a contiguous set of gas sheets that criss-cross each other 
as predicted by the equation of motion. This can be easily 
accomplished by sorting the sheets before and after 
advancing. If $L_1$ and $L_2$ are lists that contain sorted indices,
then two gas sheets are set to cross if $L_1-L_2$ is 
non-zero. Two (or more) gas sheets that are set to cross in the 
next step are joined to form a single particle whose position is 
given by the center of mass of the sheets in the set. The 
momentum and mass is assumed to be conserved. 

\begin{figure*}
   \includegraphics[width=6.5cm,clip=true,trim=0mm 0mm 0mm 0mm]{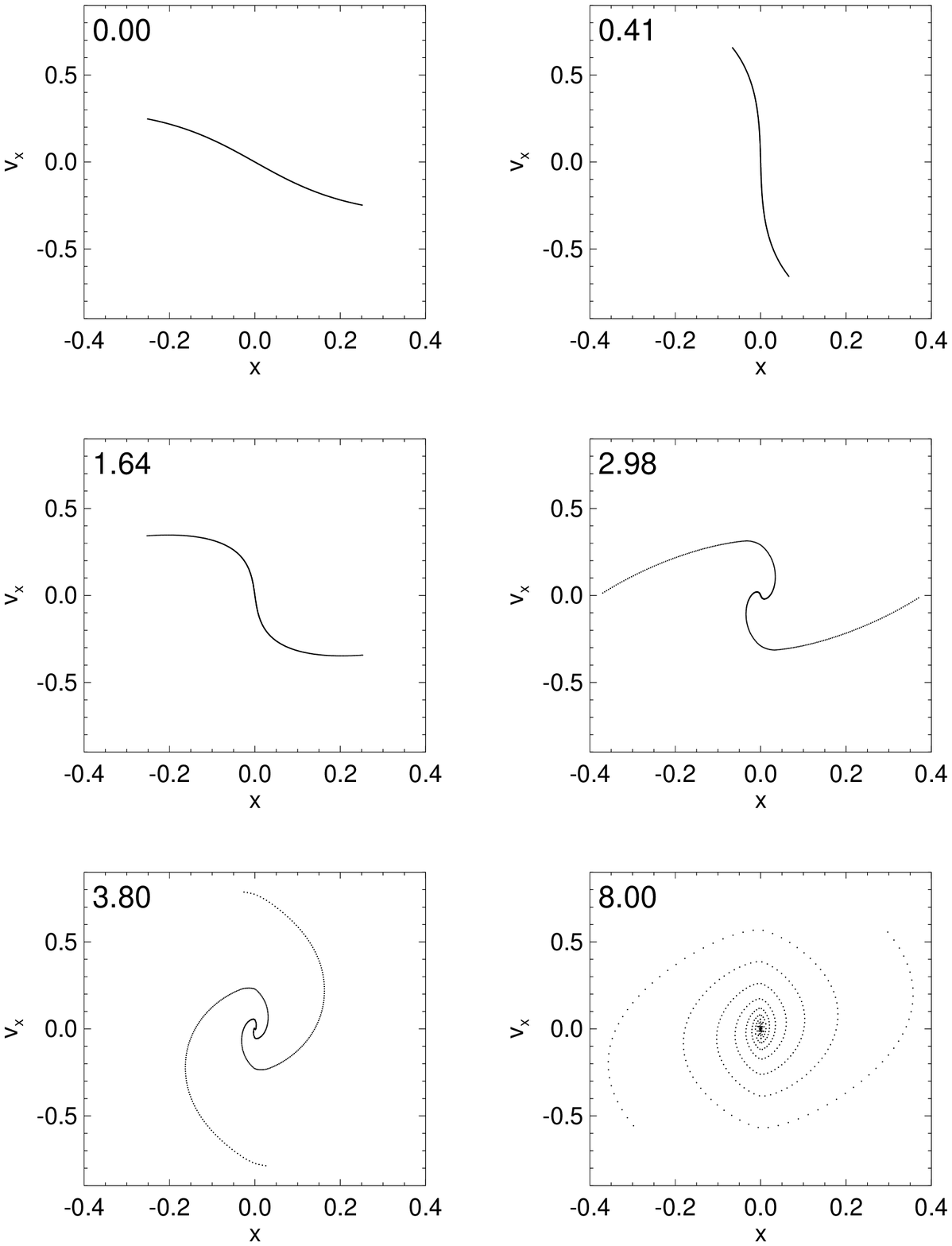}
   \includegraphics[width=6.5cm,clip=true,trim=0mm 0mm 0mm 0mm]{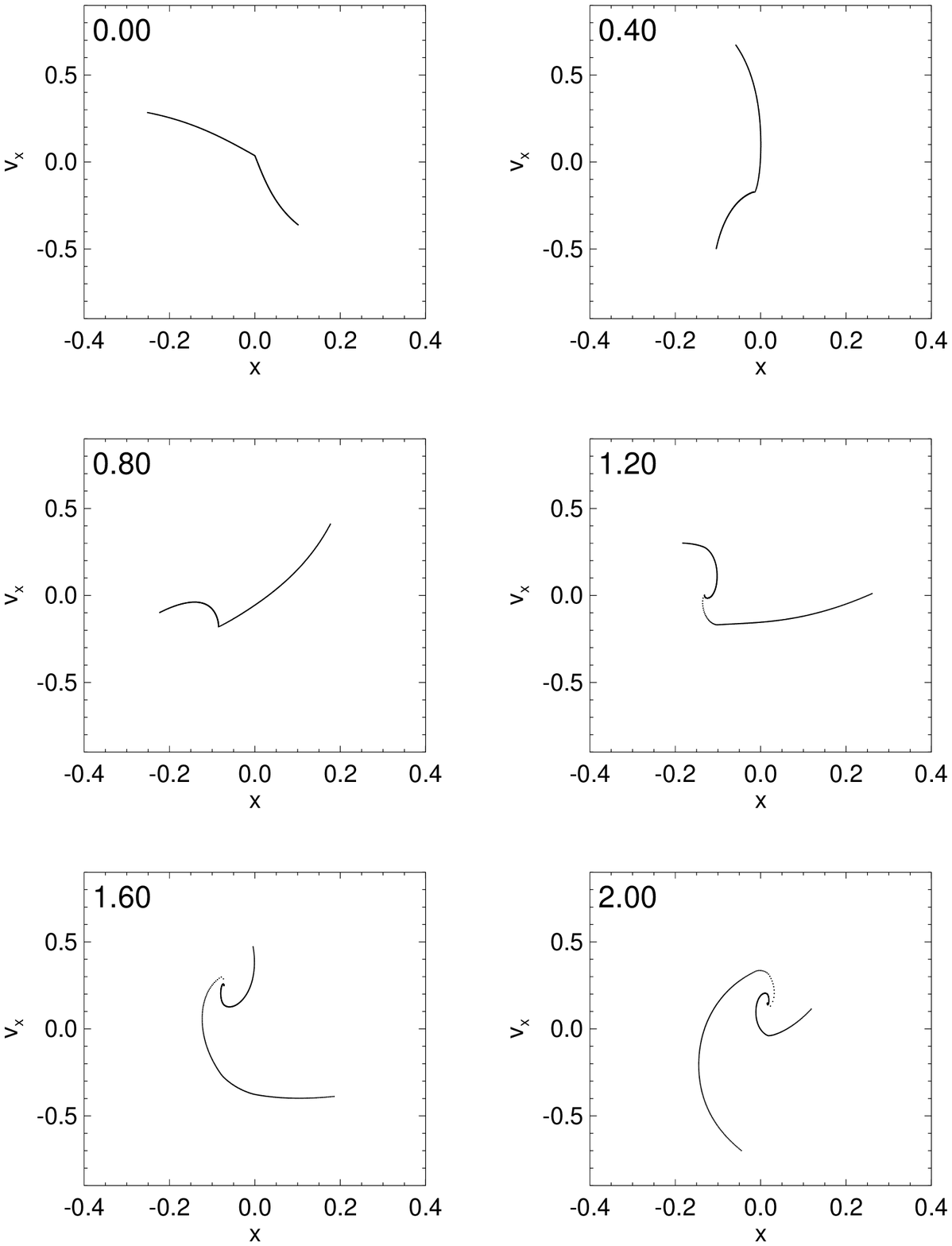}
  \caption{Evolution in phase space of parallel infinite sheets.      
     The time of the Big Bang is $-\tau$ and the units of time are in units of time $\tau$.  
     (Left) Dark matter only, symmetric perturbation. (Right) Dark matter $+$ gas, asymmetric
     perturbation.
\label{fig:evol_dm}}
\end{figure*}
In our first test, we run a simulation with dark matter. In \fig{evol_dm} (left),
we accurately recover Fig. 9.14 in BT08.
The curves are smooth and continuous in phase space, which tells 
us that the time integration is working correctly. 
We now study the evolution of sheets when both dark
matter and gas are present together. We consider both the symmetric
and asymmetric perturbations where the functional form of the perturbation is given by
\equ{perturbation}.

\begin{figure*}
   \includegraphics[width=6.5cm]{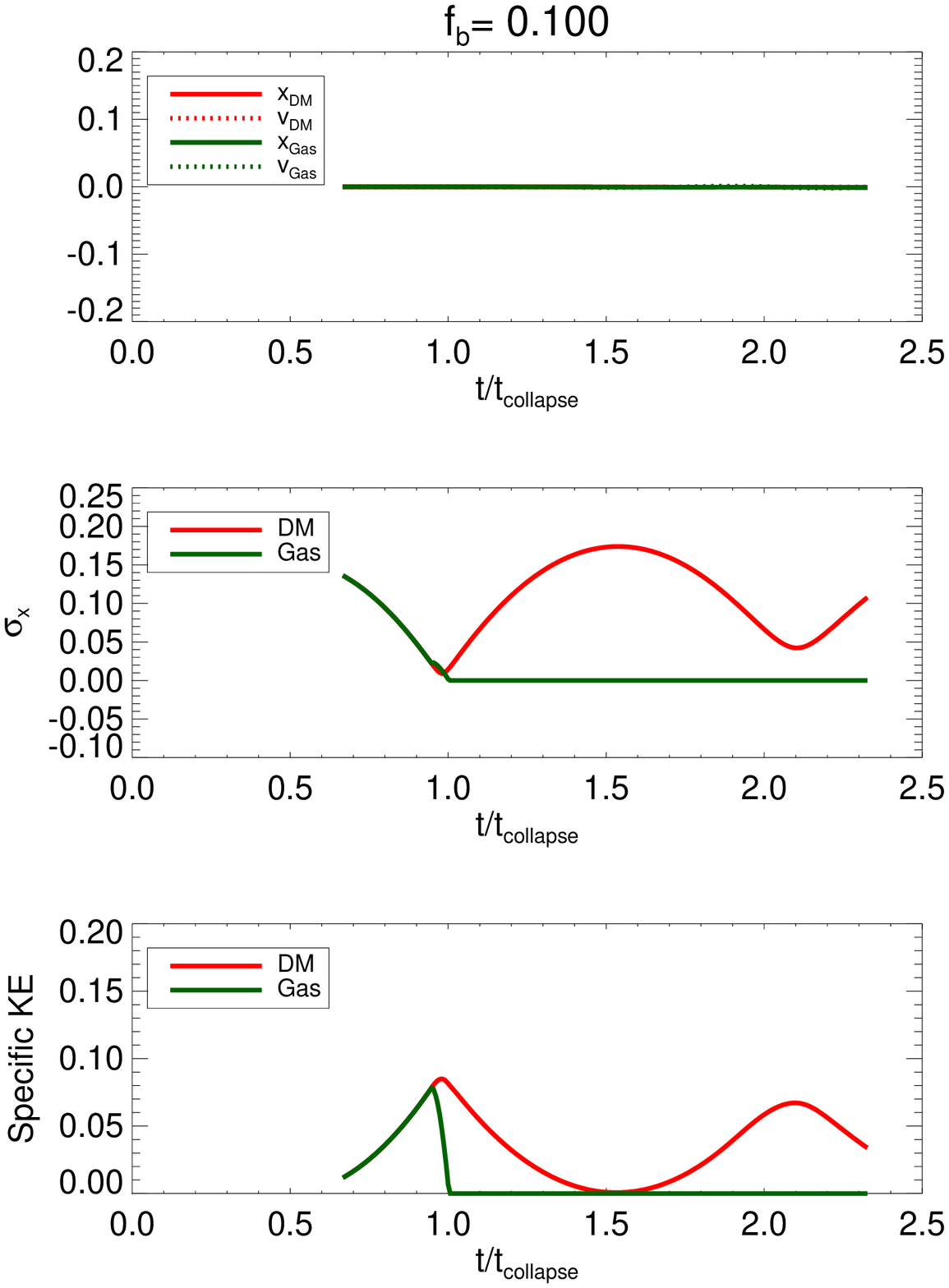}
   \includegraphics[width=6.5cm]{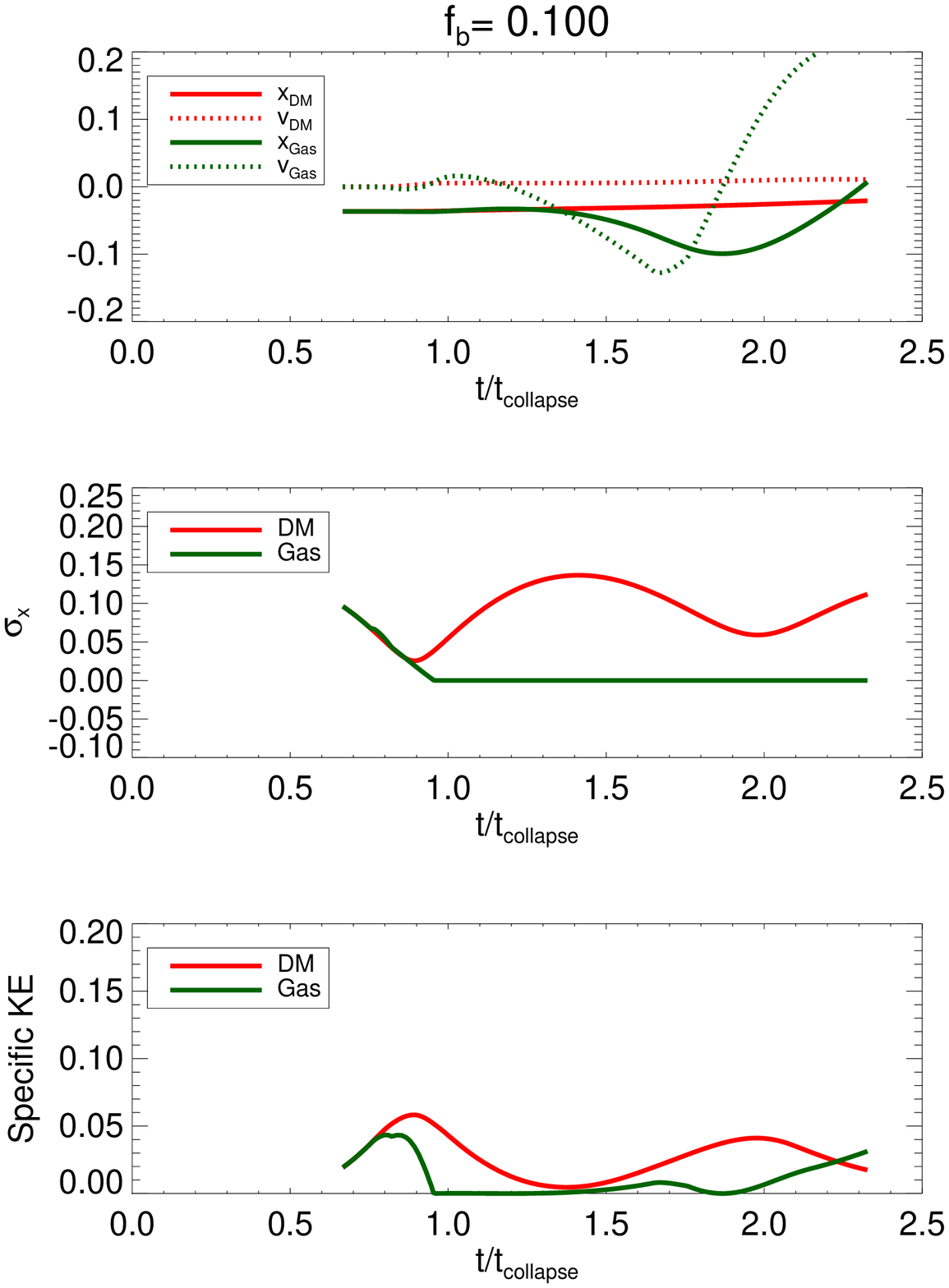}
\caption{Evolution of position and velocity of the center of mass of 
dark matter, gas and the whole system. The middle panel shows the
dispersion of sheets in space and the bottom panel shows the specific 
kinetic energy. The simulation explores the evolution of a symmetric 
perturbation (Left) and an asymmetric perturbation (Right). In the top
left figure, all of the curves are overlaid.
\label{fig:evol_f1}
}
\end{figure*}

\smallskip
\noindent{\it Symmetric perturbation.}
The first panel in \fig{evol_f1} (Left) shows the variation of the center of
mass position and center of mass velocity
with time; these are shown separately for gas ($x_{\rm gas}$,  $v_{\rm gas}$)
and dark matter ($x_{\rm DM}$,  $v_{\rm DM}$).
The middle panel shows the dispersion $\sigma_x$ or spread of the 
sheets along the $x$-axis. The dark matter sheets show oscillatory 
behaviour while the gas sheets stick after the first crossing. 
The bottom panel shows the variation in the mean kinetic energy.
The kinetic energy of the gas sheets is assumed to be lost to internal
energy within the sheets through shock heating.
We explored different baryon fractions from $f_b=0.01$ to $f_b=0.1$.
A larger value of $f_b$ increases the time required by dark matter 
to achieve the second turnaround and collapse. We observe that there is
{\it no} offset between the center of mass of the dark matter and the gas $-$
all of the lines are overlaid (see top left panel).

\smallskip
\noindent{\it Asymmetric perturbation.}
In \fig{evol_f1}, we compare our asymmetric collapse model with
the symmetric case. Clear differences are evident. The asymmetric
distribution of sheets leads to an initial (non-zero) offset in the velocity 
centroid of the gas and dark matter.
As the collapse proceeds, the gas becomes progressively separated 
from the dark matter. The dark matter starts to exert a force
on the gas and, at a later stage, gas tries to move towards the dark
matter's centre of mass. A larger value of $f_b$ increases the spatial offset and 
delays the time required by the baryons to turn around and fall towards the dark matter.
The kinematic offset between the gas and dark matter is exaggerated
here because of the artificial asymmetry built into our model. Now we turn our
attention to sheet collapse in a 3D cosmological context.

\section{Simulation with Gadget-2 in a cosmological context}

\subsection{Vacuum boundaries}
\begin{figure*}
   \includegraphics[width=6.5cm]{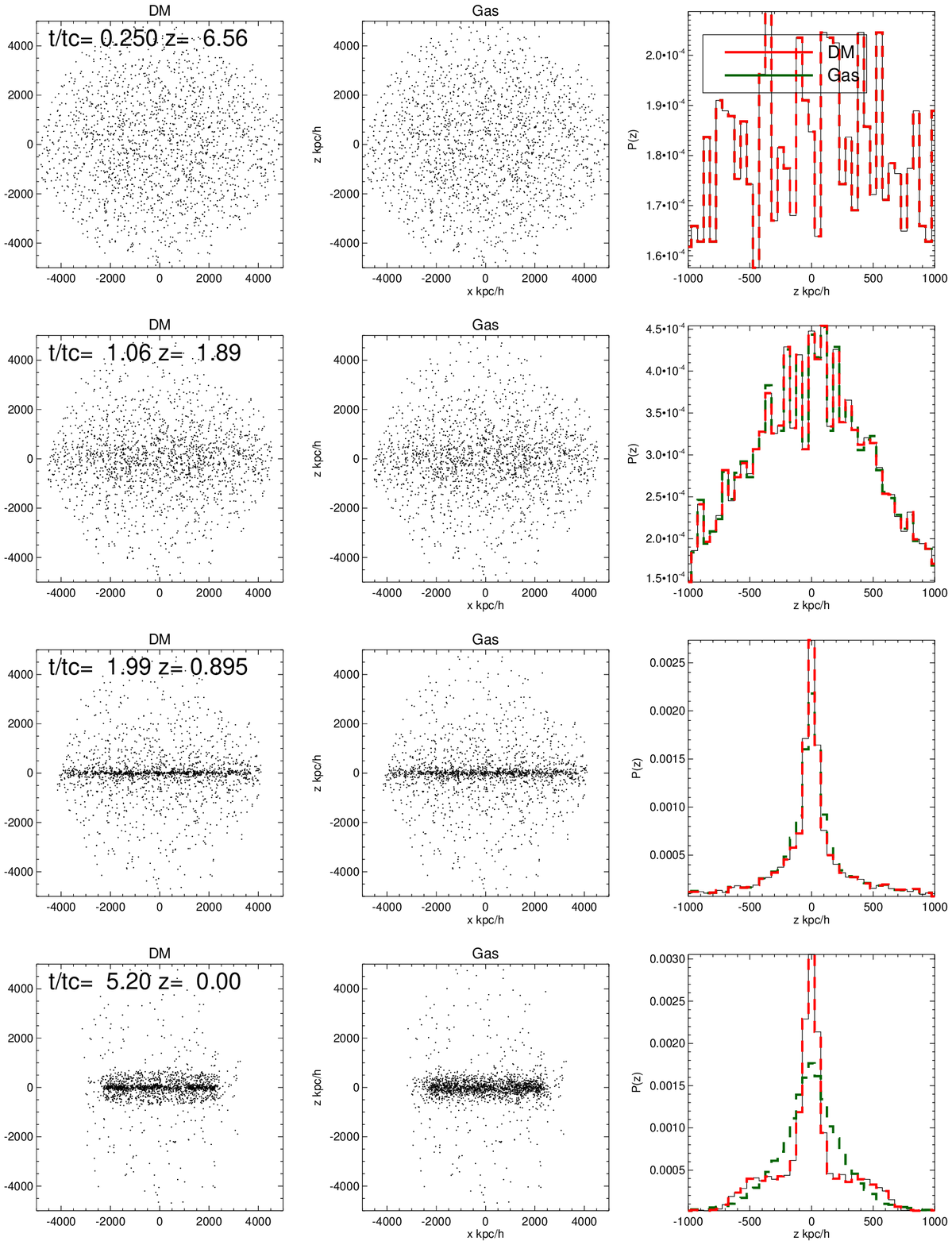}
   \includegraphics[width=6.5cm]{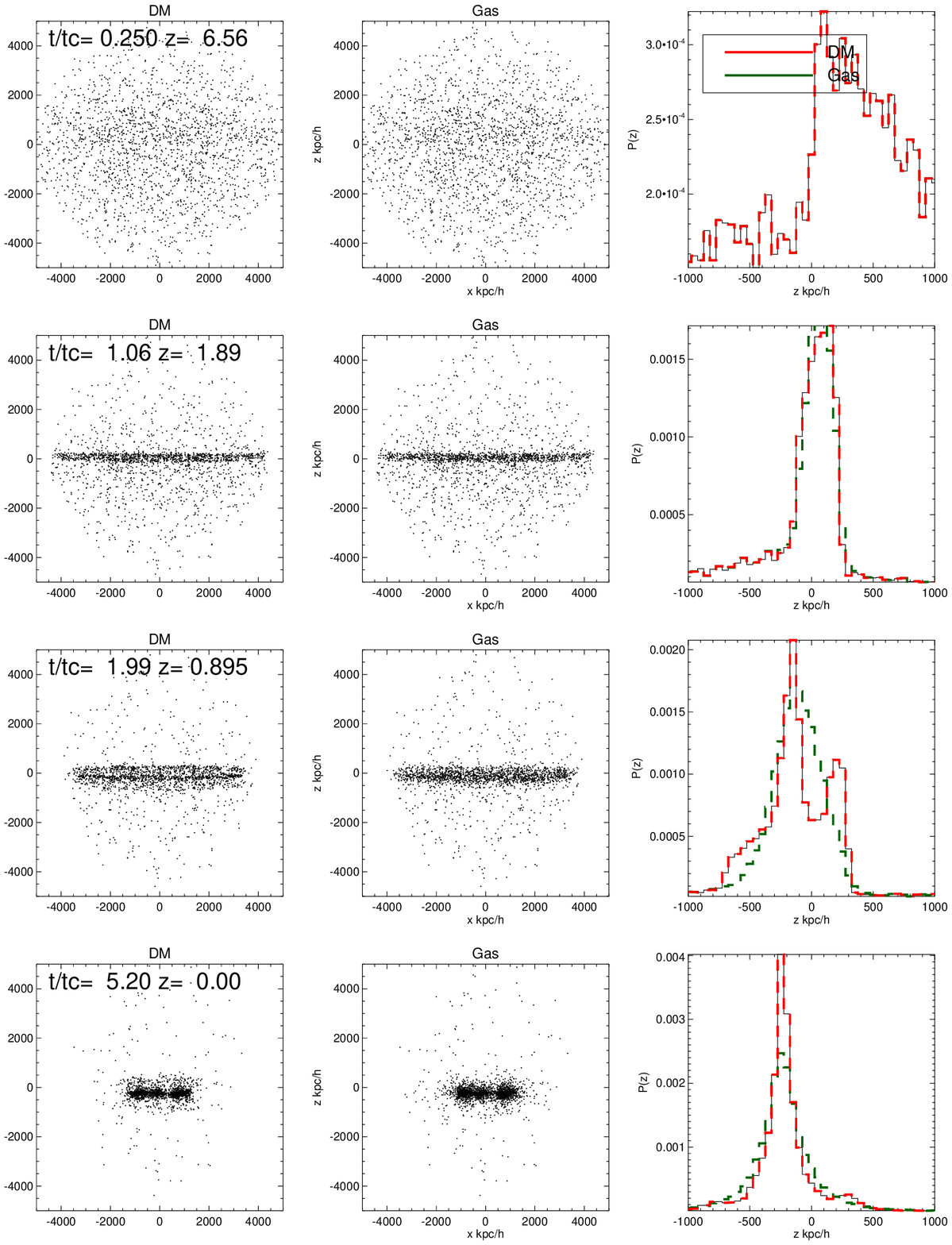}
    \caption{The distribution of particles in $x-z$ space as a function
     of time for a 10 Mpc h$^{-1}$ simulation with a symmetric cosine perturbation (left)
     and an asymmetric perturbation (right) along the $z$-direction. These are
     evolved with Gadget-2 with no gas cooling. The perturbation is
     imposed over a uniform sphere.
\label{fig:evol_f4}}
\end{figure*}

We explore a collapsing sheet where the perturbation takes the form 
\bee
\rho(z)=A\cos(2\pi z/L) .
\eee
{\it Note that the collapse is now along the $z$ axis.}
The amplitude $A$ was selected so that $\delta\rho/\rho=1$ at redshift ${\cal Z}={\cal Z}_{\rm collapse}=2$.
This perturbation was evolved using linear theory till ${\cal Z}={\cal Z}_{\rm start}=6.6$, the point 
from where the simulation starts. To set up the perturbation, the particles were initially distributed 
uniformly in a box of size $L=10$ Mpc h$^{-1}$ and a spherical region was cut
from this. We adopt a $\Omega_m=1$ and $\Omega_b=0.05$ co-moving cosmology. 
Here we use $N=32^3$ particles for the dark matter, and the same for the gas.
The displacement field corresponding to the perturbation was calculated, viz.
\bee
S(z)={{- 2\pi z}\over{LA}} \sin(2\pi z/L)
\eee
and the particles were accordingly displaced from the 
uniform distribution to generate the perturbation. The time integration was done in 
co-moving coordinates using Gadget-2. The results of the symmetric collapse (\fig{evol_f4}) 
are in good agreement with Dekel (1983).

The asymmetric case was set up by increasing the displacement field by 
a factor of two for $z>0$. In \fig{evol_f4}, we show the initial $z-v_z$ 
and density distribution of the particles.
At ${\cal Z}=0.9$, there is a bump in the density distribution 
of dark matter particles at around 400 kpc h$^{-1}$. This behaviour is not seen in the gas. 
The gas centre of mass shows a slight 
displacement with respect to dark matter similar to our idealized 1D
simulation, but the displacement is very small and is less than 
the softening parameter ($\epsilon=12.5$ kpc h$^{-1}$).
An $N=64^3$ simulation with $\epsilon=6.25$ kpc h$^{-1}$ also shows similar behaviour. 
We note that the amount of displacement in the 
1D case is much larger than for the 3D set-up in physically motivated co-moving
coordinates.

\subsection{Periodic Boundaries}
\begin{figure}
\centerline{   \includegraphics[width=6.5cm]{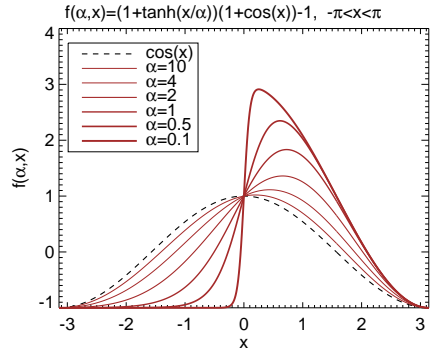}
}
 \caption{The density profile of the perturbations. A $\tanh$
     function is used to introduce asymmetry. It has well behaved derivatives 
     across the boundary and is suitable for periodic boundary conditions; 
     see $\S 3.3$ for details.
\label{fig:density2}}
\end{figure}

In earlier simulations, a plane wave perturbation was imposed on a 
uniform sphere and the asymmetry function had a discontinuity at
$z=0$. We now employ a periodic box and use a better asymmetry
function. The symmetric perturbation is a cosine function and we now
add a $\tanh$ function to make it asymmetric (see \S 3.3), such that 
\bee
\delta(z) = 
A\left[\left(1+\tanh\left(\frac{z 2\pi}{\lambda\alpha}\right)\right)\left(1+\cos\left(\frac{z
        2\pi}{\lambda}\right)\right)-1\right]
\eee
For $\alpha \gg 1$, this reduces to the plane wave form $\delta(z)=A\cos(2\pi z/\lambda)$. 
The degree of asymmetry is controlled by the shape parameter $\alpha$. The smaller the 
value of $\alpha$, the higher the asymmetry. The functional form for different $\alpha$ is 
shown in \fig{density2} (Left). The function and its first derivative 
are continuous across the boundary of the box; more details are given in \S~3.3.

We simulate 7 different types of perturbation: one is a symmetric 
cosine wave; the others use an asymmetry factor
$\alpha=100,10,1,0.5,0.25$ and $0.1$.  These were simulated with and
without cooling. A smoothing length of 12.5 kpc h$^{-1}$ was used.
$N=32^3$ particles were used once again for each constituent in the simulation and
the cosmology adopted is $\Omega_m=1$ and $\Omega_b=0.05$. 
The amplitude of the perturbation was set so as 
to make the cosine perturbation collapse at redshift ${\cal Z}_{\rm collapse}=2$. 
A starting redshift of ${\cal Z}_{\rm start}=6.6$ was used. These last two
parameters are the same as in our first Gadget-2 run with vacuum boundary
conditions. Here we only show results for the symmetric 
case and with $\alpha=1$, $0.5$ and $0.25$.

\begin{figure*}
   \includegraphics[width=6.5cm]{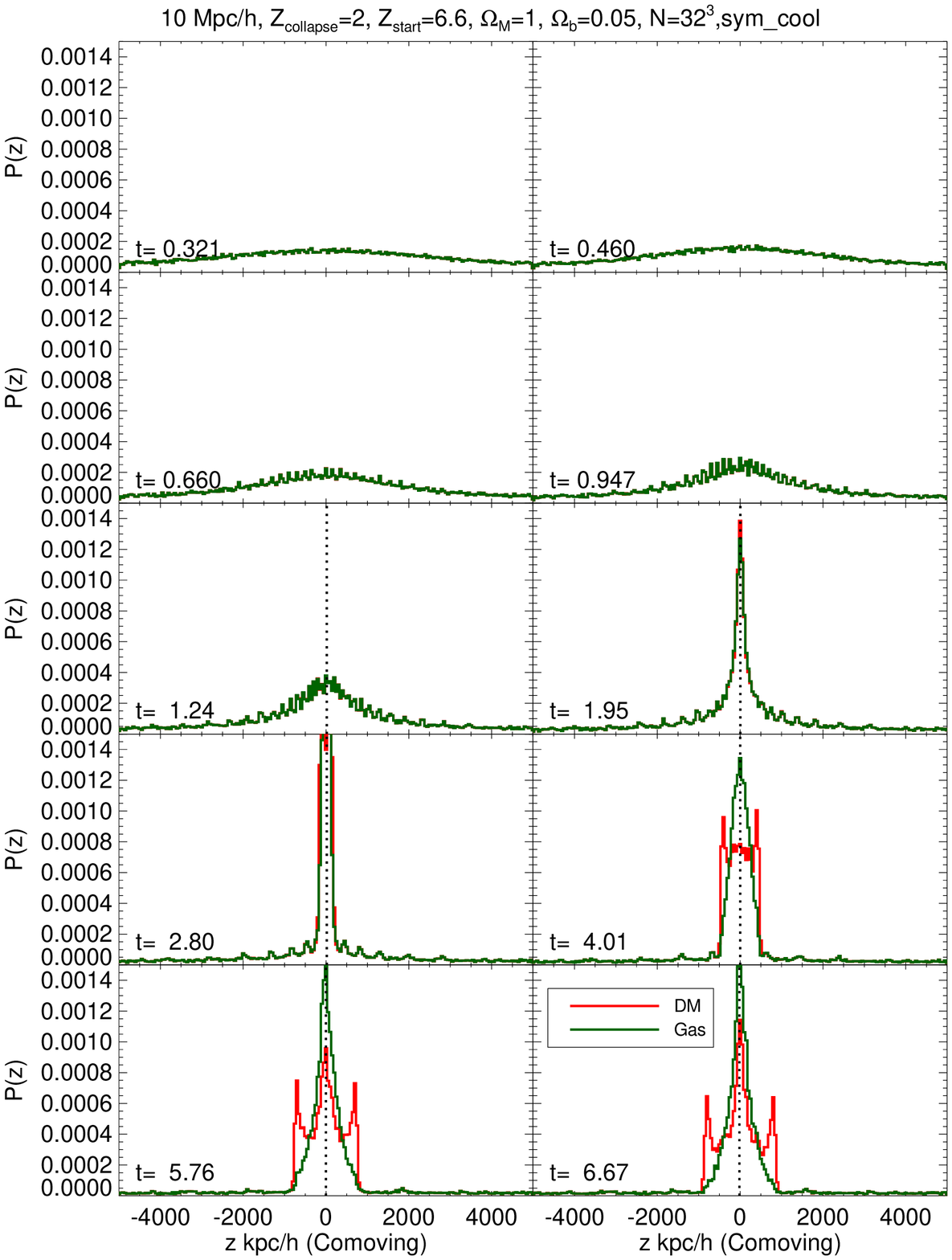}
   \includegraphics[width=6.5cm]{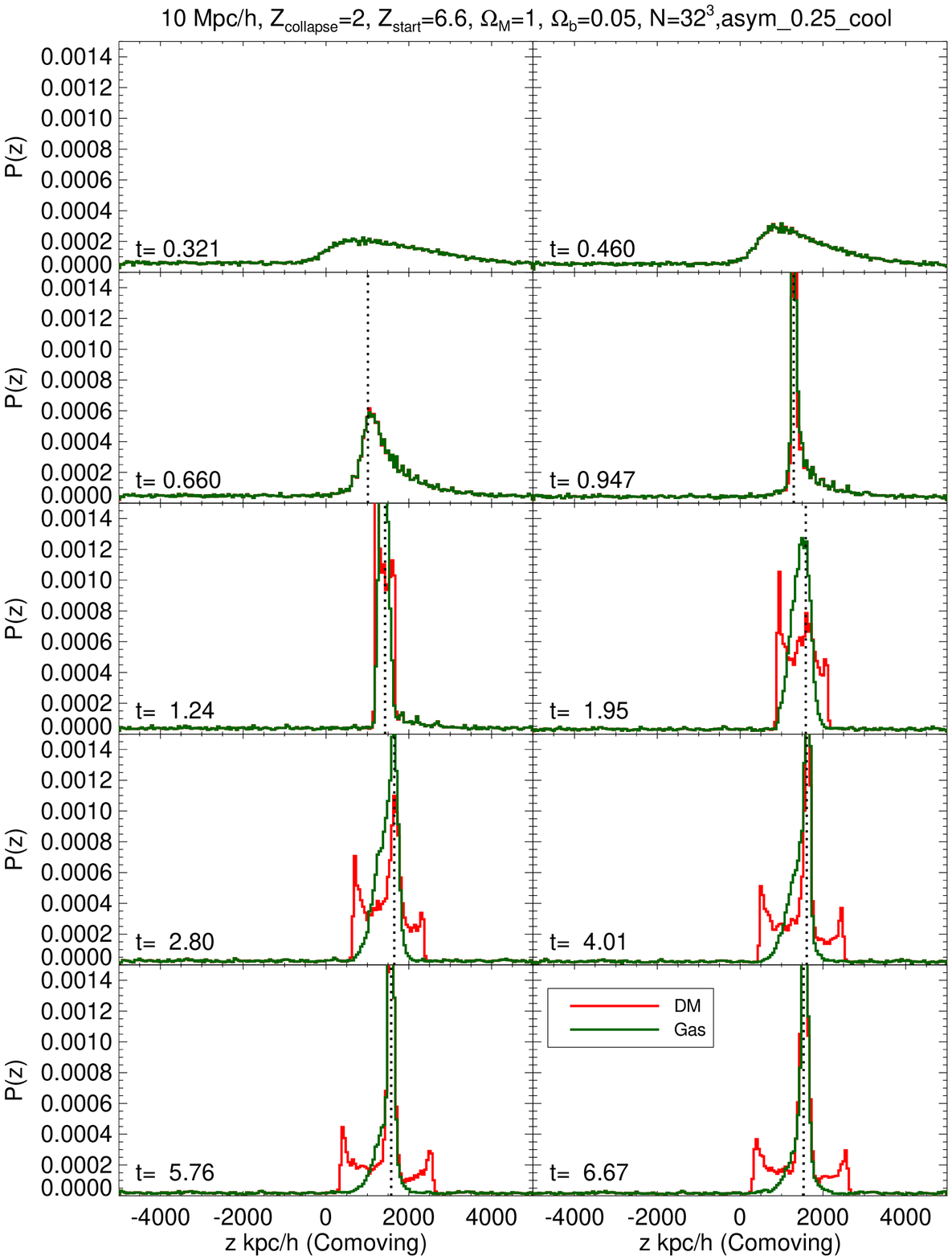}
   \caption{ The $z$ distribution of particles  as a function
     of time for a 10 Mpc h$^{-1}$, $N=32^3$  simulation. (Left) Symmetric case with cooling;
     (Right) Asymmetric perturbation $\alpha=0.25$ with cooling. The results for $\alpha \lesssim 1$
     are very similar. The vertical lines show the location of the gas
     density peak. We do not mark the peak location at very early
     times, this is because it is difficult to locate the peak 
     when the amplitude of the  perturbation is very small. 
\label{fig:evolp_f5}}
\end{figure*}

In \fig{evolp_f5}, we show the density distribution
of the dark matter and gas along the $z-$axis (in co-moving 
coordinates). The type of
perturbation and details about cooling are given on top of each
figure. To simulate gas cooling, we set an upper bound on the
temperature at $1000K$. The results with and without cooling 
are very similar. A notable feature is the occurrence of spikes 
at the edges due to the piling up of particles at turnaround after shell 
crossing. This was also noticed by Dekel (1983) 
in his simulations. The gas is not so able to criss-cross and thus 
forms a central peak. The fact that, even with cooling, the dispersion 
of the gas does not diminish is interesting. The gas appears to
expand as the DM potential becomes shallower which may be 
why cooling does not seem to have 
a great effect on the final distribution of gas.
For the asymmetric perturbation, the behaviour is similar 
except for the fact that the spikes are asymmetric.

\begin{figure}
\centerline{   \includegraphics[width=0.7\textwidth]{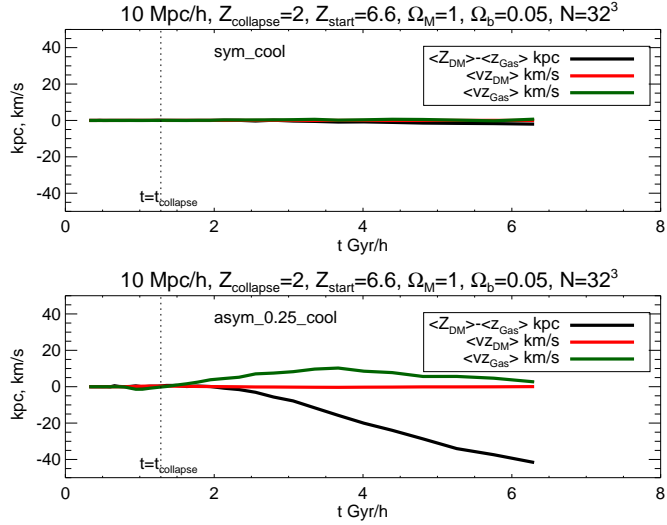}
}
  \caption{Evolution of the position and velocity  (proper) of center of mass
    of dark mater and gas for a 10 Mpc h$^{-1}$, $N=32^3$  simulation with
    periodic boundaries. (Top) Symmetric case with cooling;
     (Bottom) Asymmetric perturbation $\alpha=0.25$ with cooling. Even
   under large asymmetry, the gas to a good approximation tracks the 
   dark matter.}
\label{fig:summary_f5}
\hspace{0.05\textwidth}
\end{figure}

In the panels of \fig{summary_f5}, we plot 
as a function time the center of mass 
offset between gas and dark matter-- defined as 
$\Delta_z=\langle z_{DM} \rangle-\langle z_{Gas} \rangle$. The
velocity of center of mass of gas and dark matter is also shown on
the same panel. All quantities in these set of figures 
are in physical coordinates, i.e. physical length and physical 
peculiar velocity (without hubble flow).

For $\alpha<0.5$, i.e., large asymmetry, one 
can see that the offset $\Delta_z$ reaches to about 
$40$ kpc at redshift zero. 
The dispersion in $z$ is around 
$1000$ kpc, or an offset that is about 4\% of the 
dispersion.  Overall, the offset and velocity of the 
center of mass are quite small. Nevertheless, 
it is interesting to explore the cause of the shift.
First thing to note is that the offset occurs only 
after the collapse of the perturbation, i.e., after shell 
crossing. At this time, the dark matter
particles move past each other rapidly and hence the shape of 
the distribution is also changing rapidly 
(rapid movement of asymmetric spikes).  
The gas is less able
to criss-cross creates only a central peak and lags behind, thus
creating an offset. 

\begin{figure}
\centerline{   \includegraphics[width=0.7\textwidth]{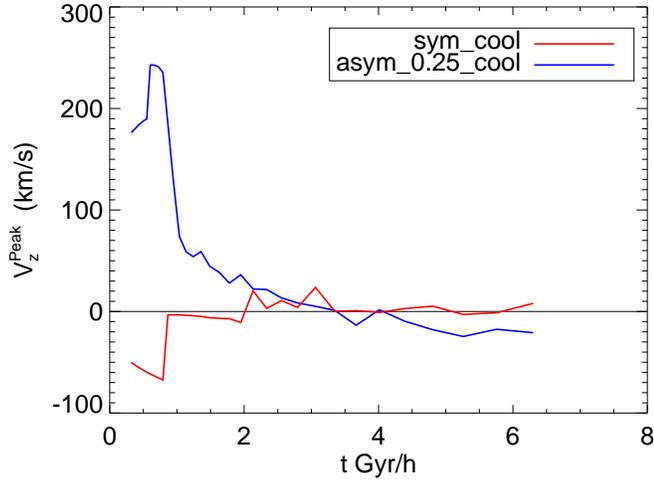}
}
  \caption{The evolution of the peculiar velocity of the peak (gas) for a 10
    Mpc, $N=32^3$  simulation with periodic boundaries. For the 
  case with the asymmetric perturbation the peak in density
  distribution is found to move with time. The peculiar velocity of
  peak is high at early times and then falls off with time.}
\label{fig:summary_f6}
\end{figure}

In \fig{evolp_f5}, the vertical lines show the location 
of the peak in comoving coordinates. If there is no peculiar 
or bulk velocity associated with peak, then the peak should 
remain stationary. For the symmetric case this is true. 
However, for asymmetric case 
the peak is not stationary. The location 
of the peak changes rapidly at earlier time, i.e., before the 
perturbation has collapsed. In \fig{summary_f6}, we  
plot the velocity of the peak, computed as the mean velocity of 
the particles in and around the peak.  For an asymmetric 
perturbation, a peculiar velocity as large as $260$ km s$^{-1}$
can be seen.

\subsection{Asymmetric Perturbation}
The asymmetric perturbation is described by the following functional form.  
\bee
f(\alpha,x) = \left(1+\tanh(x/\alpha)\right)(1+\cos{x})-1
\eee
The properties of this function are very similar to the $\cos$ function 
except that it is asymmetric. Some of the useful properties are as
follows. It is defined in range $(-\pi,\pi)$ such that
\bea
f(\alpha,-\pi) & = & f(\alpha,\pi)  =  -1, \\
\frac{df(\alpha,x)}{dx}|_{x=-\pi} & = & \frac{df(\alpha,x)}{dx}|_{x=\pi}  = 0, \\
\langle f(\alpha,x) \rangle & = & \int_{-\pi}^{\pi} f(\alpha,x) dx= 0 
\eea

An asymmetric probability distribution for a periodic box of length
$l$ with range $-l/2<x<l/2$,  is given by
\bee
p(x) = \frac{l}{4\pi^2}\left(1+A f(\alpha,2\pi x/l)\right) \nonumber
\eee
where $A$ is the amplitude of the perturbation. 
Note $p(x)>0$ only for $A\leq 1$. For $A>1$, 
this is the non-linear regime and then $p(x)$ can be negative. 
The amplitude of a cosine perturbation grows linearly 
with time. For simplicity, assuming the asymmetric perturbation to 
also behave in the same way,  
for our choice of ${\cal Z}_{\rm c}=2.0$ and ${\cal Z}_{\rm start}=6.36$, the
amplitude $A$ from the growth factor is given by 0.3968.

\section{Discussion}

With apologies to John Donne, no galaxy is an island. Most galaxies are in groups
and these accrete from the group environment which in turn accretes from the 
intergroup medium. This is a more accurate description for most galaxies than a
simple statement of that galaxies accrete from the intergalactic medium. It is only
in the last few years that modern simulations are able to show how this mechanism
operates.

Our work was motivated by new large, ongoing surveys of galaxies (Bland-Hawthorn 2014;
Bundy et al 2014) that seek to understand how the detailed properties of galaxies vary
with the local environment. Motivated by the remarkable Bullet Cluster, we 
conjectured that asymmetries would lead to a large-scale 
separation of dark matter and baryons, thereby driving a dependence of 
galaxy properties with environment, but we do not find any evidence 
for this effect. 

We do find, however, a mechanism for generating bulk flows in
the galaxy population at a level that could explain the
bulk flows of galaxies moving with respect to the cosmic microwave
background. With our Gadget-2
simulation, we show that void asymmetries in the cosmic web can
exacerbate local bulk flows of galaxies. The {\it Cosmicflows-2} survey
reveals that the Local Group 
resides in a ``local sheet'' of galaxies that
borders a ``local void'' with a diameter of about 40 Mpc
(Tully et al 2013). The void
is found to be emptying out at a rate of 16 km s$^{-1}$ Mpc$^{-1}$.
In a co-moving frame, the Local Sheet is found to be moving away from
the Local Void at $\sim 260$ km s$^{-1}$. Our model
shows how asymmetric collapse due to unbalanced voids on either side
of a developing sheet or wall can lead to a systematic movement of the
sheet, and the magnitude of the kinematic offset is (fortuitously) the same, at least
in the early stages of the sheet collapse.

Our analysis seeks to honour the memory of Y.B. Zel'dovich whose published
work continues to inspire astrophysicists around the world to the present day.
We thank the organisers for putting together such an inspiring meeting.

\acknowledgment
JBH acknowledges an ARC Australian Laureate Fellowship and SS is funded
by a University of Sydney DVC-R Fellowship. We are grateful to Merton College,
Oxford for its hospitality in the final stages of writing up this work. We acknowledge
insightful comments from T. Tepper-Garcia.


\end{document}